\documentclass[12pt]{article}
\usepackage{hyperref}
\usepackage{authblk}
\usepackage[T1]{fontenc} 
\usepackage{footnote}
\usepackage{threeparttable}
\usepackage{float}
\usepackage[utf8]{inputenc}
\usepackage[english]{babel}
 
\usepackage[nottoc]{tocbibind}

\usepackage{amsmath,amssymb}
\usepackage{graphicx}
\usepackage{caption}
\usepackage{color}
\usepackage{dcolumn}
\usepackage{bm}
\usepackage[numbers,super,comma,sort&compress]{natbib}

\newcommand{\bp}{{\mathbf p}}
\newcommand{\bq}{{\mathbf q}}
\newcommand{\br}{{\mathbf r}}

\definecolor{background-color}{gray}{0.98}

\title{Momentum space calculations of the binding energies of argon dimer}
\author[1]{Taghi Sahraeian}
\author[2,3]{M.~R.~Hadizadeh}
\affil[1]{Department of Chemistry and Biochemistry, The Ohio State University, Columbus, Ohio, 43210, USA}
\affil[2]{Institute of Nuclear and Particle Physics and Department of Physics and Astronomy, Ohio University, Athens, OH 45701, USA}
\affil[2]{College of Science and Engineering, Central State University, Wilberforce, OH 45384, USA}

\begin{document}

\maketitle

\begin{abstract}
The binding energies of argon dimer are calculated by solving the homogeneous Lippmann-Schwinger integral equation in momentum space. 
Our numerical analysis using two models of argon-argon interaction developed by Patkowski {\it et al.} confirms not only the eight argon dimer vibrational levels of the ground state of argon dimer (i.e. for $j=0$) predicted by other groups but also provides a very precise means for determining the binding energy of the ninth state which its value is a matter of discussion. Our calculations have been also extended to states with higher rotational quantum number $j$ and we have calculated the energy of all 174 bound states for both potential models. 
Our numerical results for vibrational levels of the ground state of argon dimer are in excellent agreement with other theoretical calculations and available experimental data. 
\end{abstract}

\clearpage


  \makeatletter
  \renewcommand\@biblabel[1]{#1.}
  \makeatother

\bibliographystyle{apsrev}

\renewcommand{\baselinestretch}{1.5}
\normalsize

\clearpage

\section{Introduction}
The argon dimer has been the subject of different theoretical studies and experimental research in the fields of physics and chemistry. 
The development of accurate argon-argon (Ar-Ar ) interactions has been decisive for decades and different Ar-Ar interatomic interactions have been developed by different groups, from HFDID empirical potential of Aziz \cite{Aziz_JCP99} to semi-empirical potentials of Tang and Toennies \cite{Tang_JCP118}.
Significant improvements in computational power and numerical techniques in the last decade have led to the development of highly accurate {\it ab initio} potentials advanced by Patkowski {\it et al.} \cite{Patkowski_MP103, Patkowski_JCP133} and Slavicek {\it et al.} \cite{Slavicek_JCP119}. 
Recently a new empirical potential energy function for Ar$_2$ has been also developed by Myatt {\it et al.} \cite{Myatt_MP} using a critical re-analysis of all available spectroscopic and virial coefficient data for Ar$_2$. 

In this paper, we have solved the Lippmann-Schwinger equation in momentum space to calculate the binding energies of argon dimer, using two models of Ar-Ar interaction of Refs. \cite{Patkowski_MP103, Patkowski_JCP133}. Both potential models support nine bound states for $j=0$; however, defining the shallowest state which is very close to dissociation is numerically challenging. This has been discussed in details. 
Moreover, we extend the solution of Lippmann-Schwinger equation to higher partial wave channels and present the numerical results for 174 vibrational and rotational energy levels of argon dimer.
 
\section{Lippmann-Schwinger integral equation in momentum space}

The nonrelativistic bound state of two particles with masses $m_1$ and $m_2$, and the relative momentum $\bp$, interacting by an arbitrary central force $V$ can be described by homogeneous Lippmann-Schwinger equation
\begin{equation}
\label{eq.LS_operator}
\psi = G_0 V \psi ,
\end{equation}
where $\psi$ is two-body (2B) wave function and $G_0= (E-\frac{p^2}{2\mu})^{-1}$ is free propagator.
$E=(m_1+m_2 - m_d) \cdot c^2$ is the binding energy of 2B system ($m_d$ is the mass of dimer) and $\mu= \frac{m_1 m_2}{m_1 +m_2}$ is the reduced mass of 2B system.
In a partial wave representation Eq. \ref{eq.LS_operator} can be presented in momentum space as the following integral equation \cite{Hadizadeh_PLB753, Hadizadeh_AIP}
\begin{equation}
\label{eq.LS}
\psi_{vj} (p) = \frac{1}{E-\frac{p^2}{2 \mu}} \, \int_{0} ^{\infty} dp' \, p'^2 \, V_j (p,p') \, \psi_{vj} (p'),
\end{equation}
where $v$ and $j$ are the vibrational and rotational quantum numbers, respectively.
The input of the integral equation \ref{eq.LS} is the two-body interaction $V_j (p,p')$ which depends on the magnitude of the initial and final 2B relative momenta $p$ and $p'$. $V_j (p,p')$ is the projection of the interaction $V(\bp,\bp') \equiv V(p,p',x)$ in the partial wave channel $j$
\begin{equation}
\label{eq.Vl}
V_j (p,p') = 2 \pi \int_{-1} ^{+1} dx \, P_j(x) \, V(p,p',x).
\end{equation}
The matrix elements of the interaction in momentum space $V(p,p',x)$ can be obtained by Fourier transformation of 2B interaction in configuration space $V(r)$, which depends on the relative distance $r$, as
\begin{eqnarray}
\label{eq.V-Fourier}
 V(p,p',x) &=& \frac{1}{(2\pi)^3} \int d^3r \, e^{i\bq \cdot \br} \, V(r) ; \quad \bq = \bp - \bp' \cr
 &=& \frac{1}{2\pi^2 q}  \int_{0}^{\infty} dr \, r \, \sin(qr)  \, V(r).
\end{eqnarray}
$p$ and $p'$ are the magnitudes of initial and final 2B relative momentum vectors, $x$ is the angle between them, and $q=\sqrt{p^2+p'^2-2pp'x}$ is the magnitude of the momentum transfer.

\section{Argon-argon interatomic potentials} \label{sec.potentials}

In this study for Ar-Ar interatomic interaction, we have used the fitted analytical potential functions of Refs. \cite{Patkowski_MP103,Patkowski_JCP133} with the following general form
\begin{eqnarray}
\label{Vr}
  V(r) =  \left \{ 
  \begin{array}{ll}
\left(A+BR+C/R+DR^2 +ER^3 \right) e^{-\alpha R - \beta R^2}  - \sum\limits_{n=3}^8 C_{2n}\,f_{2n}(bR)\,R^{-2n}, & R \le D \cr\cr
 \left(A'+B'R+C'/R+D'R^2 \right) e^{-\alpha' R - \beta' R^2} ,  & R> D 
 \end{array}
 \right. 
\end{eqnarray}
where $R=\frac{r}{r_m}$, and $f_{2n}(x)$ are the Tang-Toennies damping functions
\begin{eqnarray}
f_{2n}(x)=1-e^{-x}\,\sum\limits_{k=0}^{2n} \displaystyle\frac{x^k}{k!}.
\end{eqnarray}

In our calculations, we have used the potential parameters of Refs. \cite{Patkowski_MP103,Patkowski_JCP133} which are given in Table \ref{Table.Ar-parameters}.
The potential models I and II predict a minimum at $3.7673066$ \AA ~and $3.762382$ \AA ~with a depth of $99.269155\, \text{cm}^{-1}$ and $99.351070\, \text{cm}^{-1}$, respectively.
In Fig. \ref{Fig.V} we have shown both potential models I and II of Ar-Ar interaction in configuration space as a function of the interatomic distance $r$. We have also presented few examples of the matrix elements of the potential model I in momentum spaces, as a function of the relative momenta $p$ and $p'$, for the rotational quantum numbers $j = 0, 1$ and 2.

\section{Numerical Results and Discussion} \label{results}

The first step toward the numerical solution of the integral equations (\ref{eq.LS}), (\ref{eq.Vl}), and (\ref{eq.V-Fourier}) is discretization of the continuous momentum, angle, and configuration variables and to this aim, we have used Gauss-Legendre quadratures.
The momentum integration interval $[0,\infty)$ is covered by a combination of hyperbolic and linear transformations of Gauss-Legendre points from the interval $[-1,+1]$ to the intervals $[0,p_1]  \cup [p_1,p_2]  \cup [p_2,p_3]$ as
 \begin{eqnarray}
 \label{eq.mapping}
p_{\text{hyperbolic}} = \frac{1+x}{\frac{1}{p_1} + (\frac{2}{p_2}-\frac{1}{p_1} )\, x }, \quad p_{\text{linear}}=\frac{p_3-p_2}{2}\, x + \frac{p_3+p_2}{2}.
\end{eqnarray}
The typical values for $p_1, \, p_2$, and $p_3$ in our calculations are $5, \, 10$, and $100$ \AA$^{-1}$.
For the angle integration $x$ in Eq. (\ref{eq.Vl}) as well as configuration integration $r$ of the Fourier transformation of the potential in Eq. (\ref{eq.V-Fourier}), a linear transformation is used. 
The configuration integration domain is transformed to the interval $[0,r^{max}]$, where we have used $r^{max} = 600 $ \AA. 
The number of grid points for momentum, angle, and configuration variables are $300$, $100$ and $4000$, respectively.

The integral equation (\ref{eq.LS}) can be written schematically as eigenvalue equation $\lambda \, \psi = K(E)\, \psi$, where the physical dimer binding energies are corresponding to the eigenvalue $\lambda=1$. The eigenvalue equation is solved by direct method. For calculation of argon dimer binding energies we have solved the integral equation (\ref{eq.LS}) by searching in a wide range of energies in the region $-100\, \text{cm}^{-1}  \le E \le 0\, \text{cm}^{-1}$ and we have extracted the physical bound states for $\lambda=1$ with a relative error of $10^{-10}$.

Our numerical results for all 174 vibrational and rotational energy levels of argon dimer using potential models I and II are listed in Tables \ref{Table.BE_spd1} and \ref{Table.BE_spd2}, respectively. 
In Tables \ref{Table.BE_s} and \ref{Table.BE_pd}, however, our vibrational levels of argon dimer, for the rotational states $j=0, 1, 2$, are compared with those of previous studies by Tennyson {\it et al.} using R-matrix method \cite{Tennyson_FD, Rivlin_private} and also with the findings of Ref. \cite{Patkowski_private}. 
The comparisons indicate that our results for both potential models I and II are in an excellent agreement with the results of other groups with a relative percentage difference estimated to be at most $0.15\,\%$, $0.16\,\%$, and $0.19\,\%$ for $j=0$, $j=1$, and $j=2$, respectively. 
It should be pointed out that the argon mass used in our calculations and also in Refs. \cite{Tennyson_FD, Rivlin_private} corresponds to pure $^{40}$Ar mass of $39.9623831225$ amu, whereas the results of Ref. \cite{Patkowski_private} are obtained by $^{40}$Ar mass of $39.962384$ amu.

In Table \ref{Table.BE_s}, we have also listed the experimental data of Herman {\it et al.} for the binding energies of six of the nine vibrational levels for $j=0$, determined by high-resolution VUV emission from three excited electronic states of Ar$_2$.

As it is shown in Table \ref{Table.BE9}, there is an evidence in Ref.  \cite{Barletta_NJP12} suggesting the potential model I should support a ninth vibrational level for argon dimer ground state for $j=0$ with a very small binding energy in the microKelvin range. Our numerical analysis confirms the existence of this extremely shallow state with a binding energy of $-1.24070\, \mu$K which is in good agreement with the reported energy of $-0.74$ $\mu$K in Ref. \cite{Barletta_NJP12}.
Although the R-matrix method \cite{Tennyson_FD} has been able to detect only eight of the nine vibrational weakly bound states, some recent improvements in this method \cite{Rivlin_private} have provided the possibility to generate the ninth vibrational state for both potential models I and II, however, at this point, the exact value of the energy level of the ninth state is unstable and that is why it is not listed in Table \ref{Table.BE_s}.

We should point out that Myatt {\it et al.} have recently performed a full rotational treatment using a new empirical potential energy function for Ar$_2$ and have reported 173 vibrational and rotational energy levels \cite{Myatt_MP}. In their numerical analysis, the employed potential energy supports only eight, rather than nine vibrational levels. 
For the numerical implementation, the authors have used a computer code called "LEVEL", developed by R. J. Le Roy to solve the radial Schr\"odinger equation for diatomic systems and calculate the eigenvalues of the bound and quasibound levels of any smooth one-dimensional or radial potential \cite{LEROY2017167}. 
The numerical analysis of Ref. \cite{Rivlin_private} confirms that not only the R-matrix method of Ref. \cite{Tennyson_FD}, but also LEVEL code has been unable to detect the ninth vibrational weakly bound state, whereas our momentum space method provides a superior solution to the Schr\"odinger equation with the capability to achieve the extremely weakly bound states. 
By having dimer binding energies one can calculate the argon dimer wave functions using Eq. (\ref{eq.LS}). In Fig. \ref{Fig.WF} we have shown dimer wave functions for all nine vibrational states of the ground state of argon dimer, for $j=0$, obtained from the potential model I, as a function of the relative momentum $p$. The dimer wave functions are normalized as $\int_0^{\infty} dp\, p^2 \, \psi^2_{vj}(p) = 1.$ 
As we expect, the structure of dimer wave function is expanded to higher momentum for lower bound levels, whereas for the higher bound levels the wave functions are more compact. As it is shown in Fig. \ref{Fig.WF}, for the ninth vibrational state for $j=0$ which is weakly bound and very close to dissociation, the wave function is significant at very small values of the relative momentum $p$. 
Our numerical analysis indicates that the binding energy of this weakly bound state is quite sensitive to the distribution of the grid points close to zero momentum. In order to be able to reasonably identify this state, it is crucial to consider a hyperbolic mapping with enough number of mesh points for the relative momentum $p$ in the domain $10^{-4} ~\text{\AA} \le p \le10^{-1}~\text{\AA}$. 
In Fig. \ref{Fig.density}, we have shown few examples of the vibrational radial probability densities $p^2|\psi_{vj}(\bp)|^2$ of argon dimer ground state for $j=0$, calculated for potential model I. 
We have also shown few examples of the rotational probability densities $|\psi_{vj}(\bp)|^2$ for $v=2$ and $j=0$, $j=1$, and $j=2$. As we can see the argon dimer radial probability densities for higher states have been squeezed to smaller momenta which as we expect leads to larger expectation values for the relative distance between Ar-Ar pair. 

\section{Conclusion}
We have presented the numerical solution of Lippmann-Schwinger integral equation in momentum space to calculate argon dimer binding energies. We have provided full rotational and vibrational energy levels of argon dimer using two models of argon-argon interaction.
We have shown in this paper that the solution of Lippmann-Schwinger integral equation in momentum space has the advantages of catching the weakly bound states and numerically less challenging than the solution of Schr\"odinger equation in configuration space, where the two-body wave function has been expanded to a very large relative distance.
Instead, in momentum space, the behavior of the wave function is reversely and for shallow bound states, the wave function is squeezed to a very small momentum which can be numerically treated much more efficiently. 
Our numerical results for the vibrational levels of the ground state of argon dimer are in excellent agreement with other theoretical predictions, especially with those reported by Patkowski {\it et al.} and also by recent results obtained by R-matrix method. 
The comparison of our numerical results for dimer binding energies with other theoretical results and experimental data indicates that our method is technically feasible, reliable, and a highly efficient method to determine the dimer binding energies.
In the next step, we are going to extend a direct integration method called "three-dimensional" approach \cite{Hadizadeh_FBS40,Hadizadeh_PRC83}, which has been successfully applied to nuclear bound and scattering systems and avoids the traditional partial wave representation and its complexity, to atomic three- and four-body bound states.

\subsection*{\sffamily \large ACKNOWLEDGMENTS}
The authors thank Konrad Patkowski for providing dimer binding energy values of the potential models I and II, and also to Tom Rivlin for helpful discussions and for providing dimer binding energies obtained by R-matrix method. 
This work is performed under the auspices of the National Science Foundation under Contract No. NSF-HRD-1436702 with Central State University. M. R. H. acknowledges the partial support from the Institute of Nuclear and Particle Physics at Ohio University. 
\clearpage



\bibliography{references}


\clearpage

\begin{figure}
\caption{\label{Fig.V} Upper panel: the Ar-Ar interaction as a function of the interatomic distance $r$ for the potential models I and II. 
Lower panel: the matrix elements of the Ar-Ar interaction of potential model I in momentum space as a function of the relative momenta $p$ and $p'$, calculated for the rotational quantum numbers $j=0$ (left panel), $j=1$ (middle panel), and $j=2$ (right panel) in units of K \AA$^3$. }
\end{figure}

\begin{figure}
\caption{\label{Fig.WF} Upper panel: the argon dimer wave function $\psi_{vj}(p)$ as a function of the relative momentum $p$, for nine vibrational levels of the ground state of argon dimer, i.e. $j=0$, calculated for the potential model I, given in Table \ref{Table.BE_s}. Lower panel: the absolute value of the ninth vibrational dimer wave function for $j=0$ in a logarithmic scale.}
\end{figure}

\begin{figure}
\caption{\label{Fig.density} Upper panel: the radial probabilities $p^2|\psi_{vj}(\bp)|^2$ for all nine vibrational states of argon dimer ground state, i.e. $j=0$, calculated for the potential model I.
 Lower panel: the $x-z$ cross section of the probability densities $|\psi_{vj}(\bp)|^2$ for the vibrational quantum number $v=3$ and the rotational quantum numbers $j=0$ ($s-$wave), $j=1$ ($p-$wave), and $j=2$ ($d-$wave).}
\end{figure}



\clearpage

\begin{center}
$
\begin{array}{cc}
\includegraphics[width=.45\textwidth]{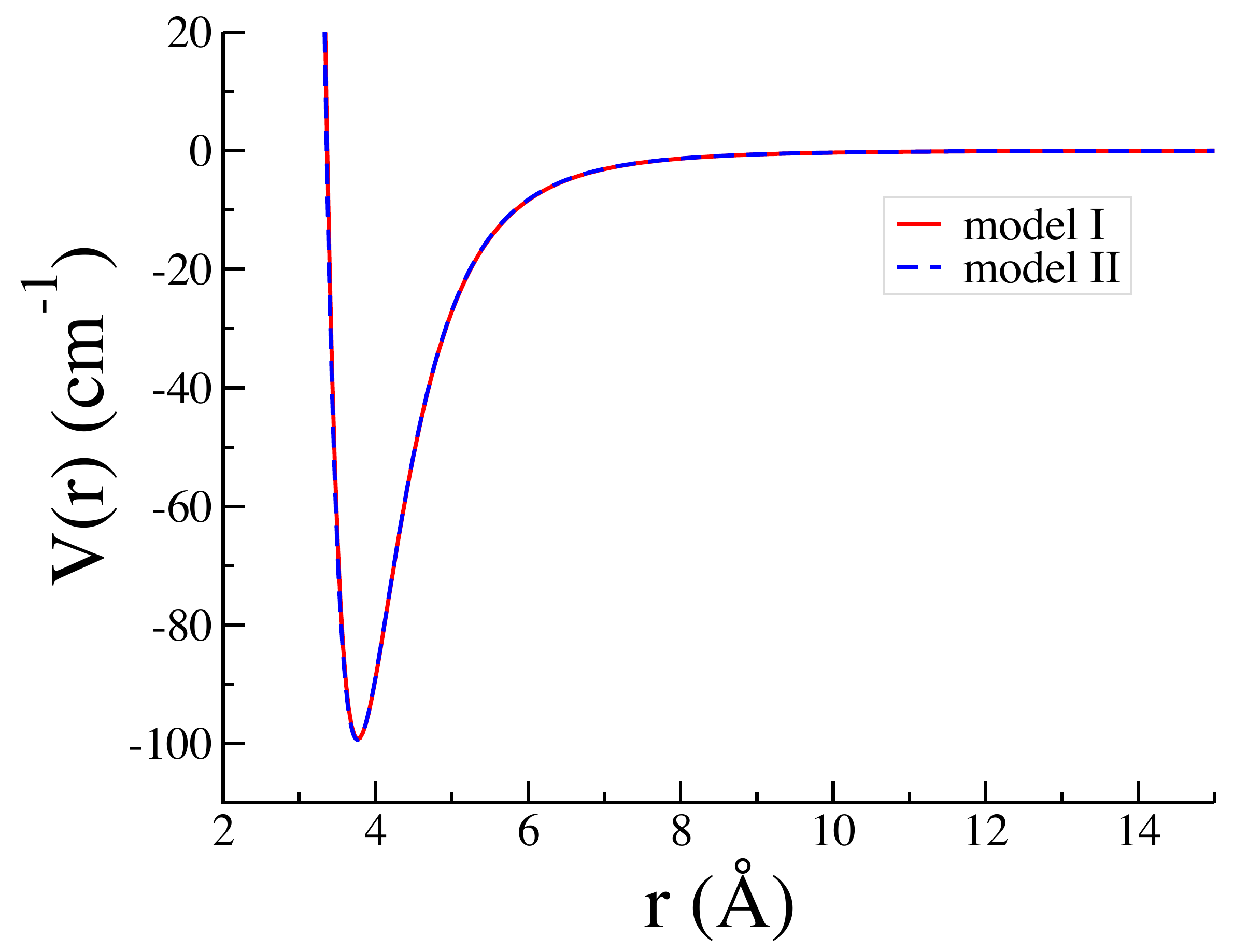}   \\ 
 \includegraphics[width=.9\textwidth]{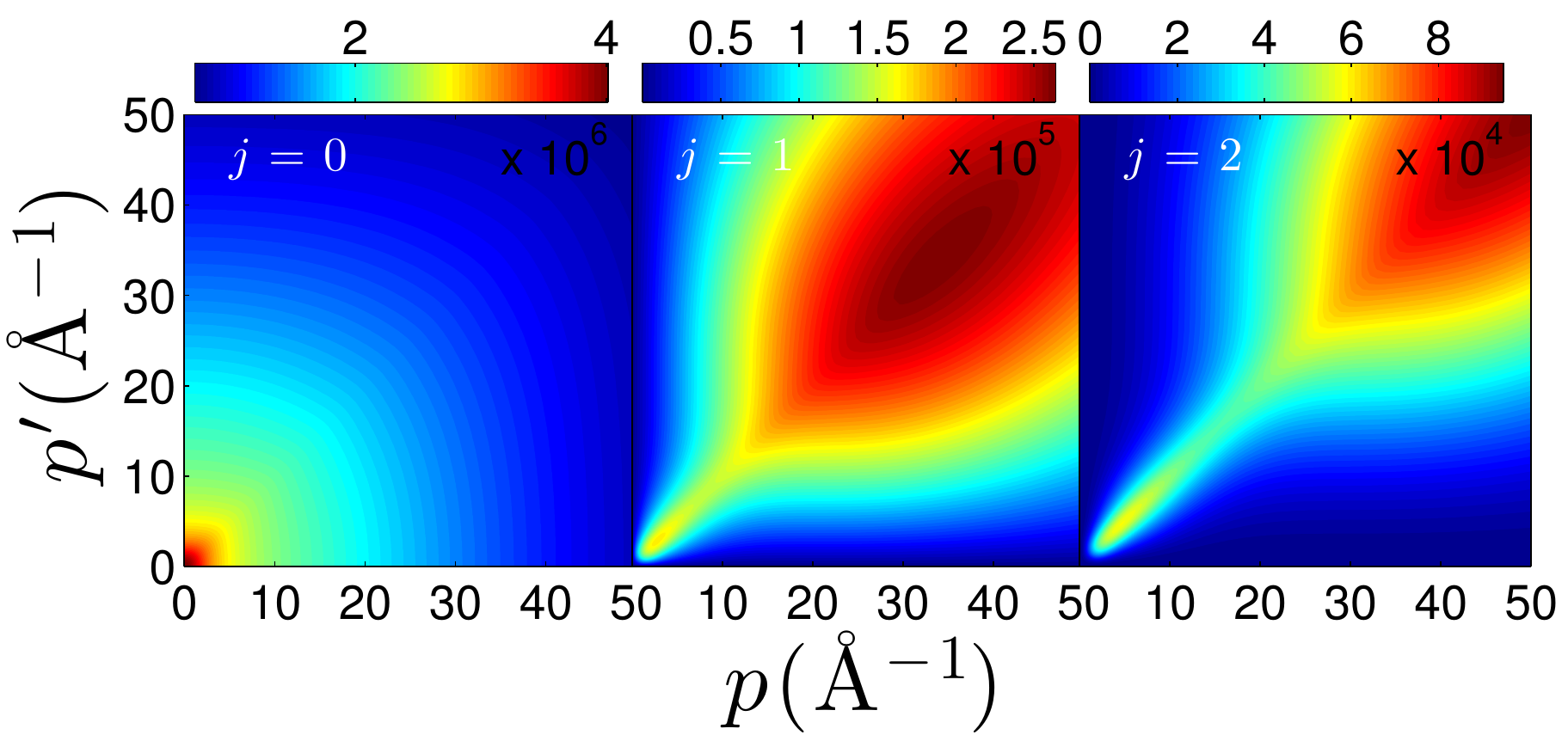}   \\ 
\end{array}  
$
\end{center}
\vspace{0.25in}
\hspace*{3in}
{\Large
\begin{minipage}[t]{3in}
\baselineskip = .5\baselineskip
Figure 1 \\
Taghi Sahraeian, M.~R.~Hadizadeh  \\
Int. J.\ Quant.\ Chem.
\end{minipage}
}

\begin{center}
$
\begin{array}{cc}
\includegraphics[width=.6\textwidth]{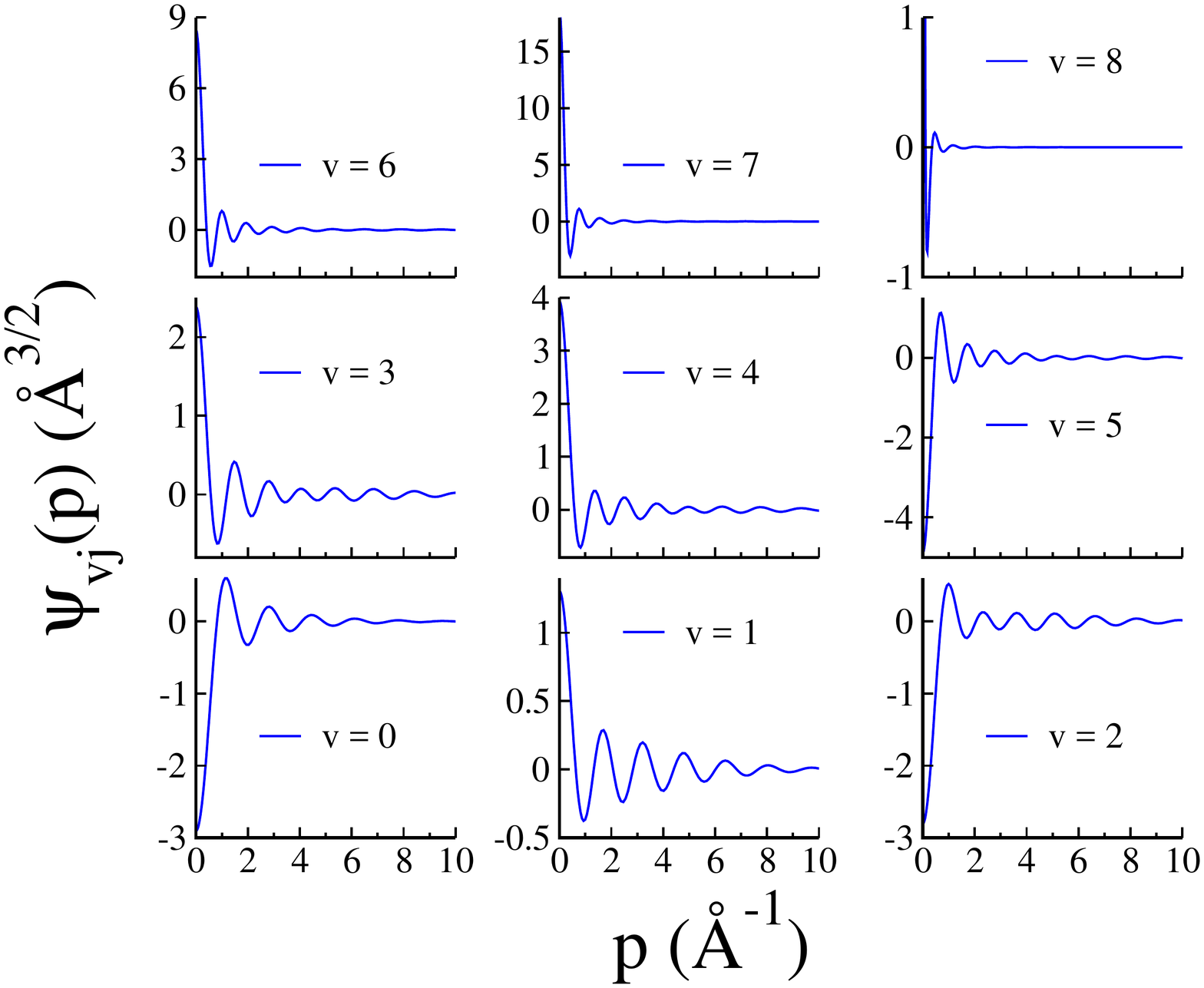}   
\\
  \includegraphics[width=.5\textwidth]{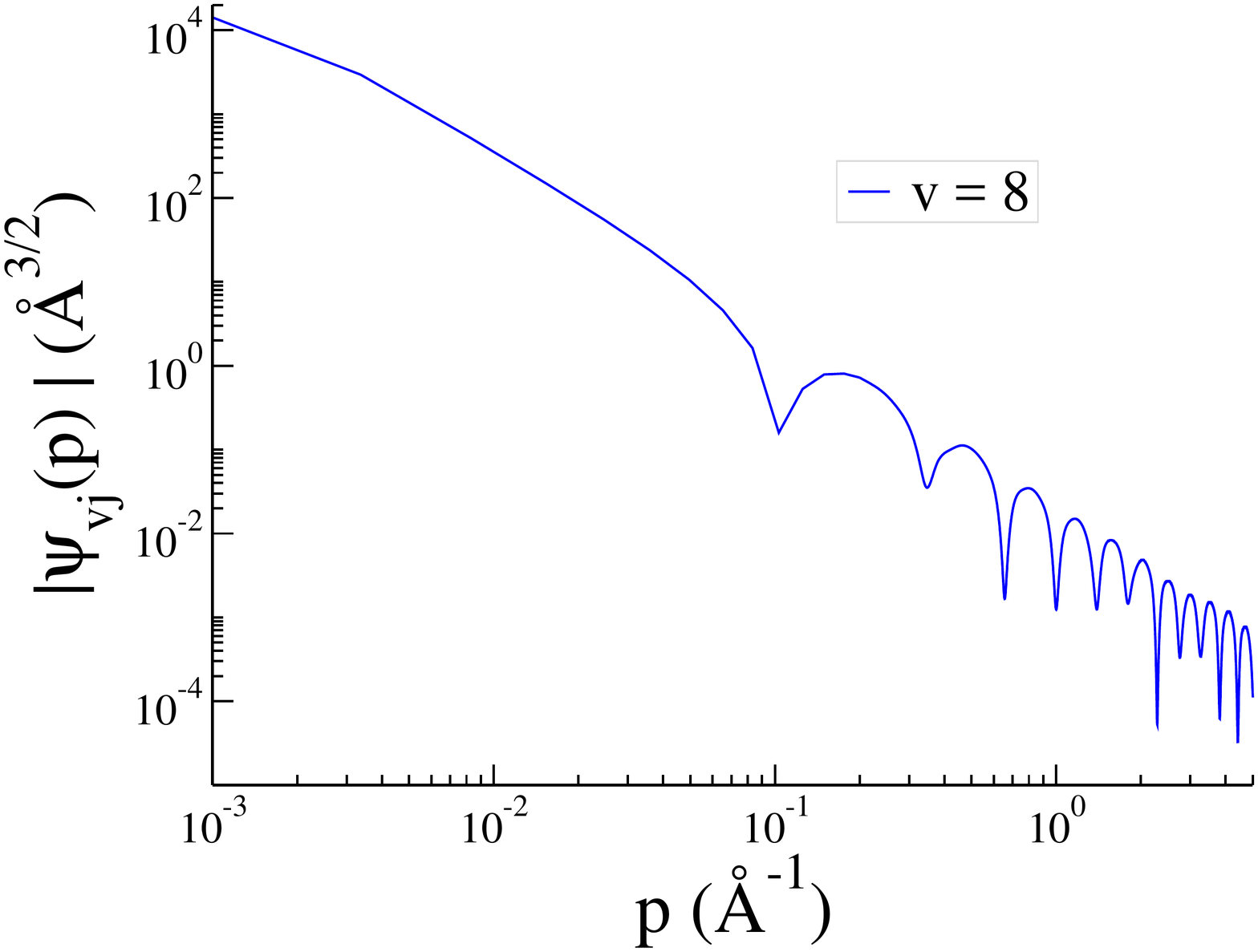}   \\ 
\end{array}  
$
\end{center}
\vspace{0.25in}
\hspace*{3in}
{\Large
\begin{minipage}[t]{3in}
\baselineskip = .5\baselineskip
Figure 2 \\
Taghi Sahraeian, M.~R.~Hadizadeh  \\
Int. J.\ Quant.\ Chem.
\end{minipage}
}

\begin{center}
$
\begin{array}{cc}
\includegraphics[width=.6\textwidth]{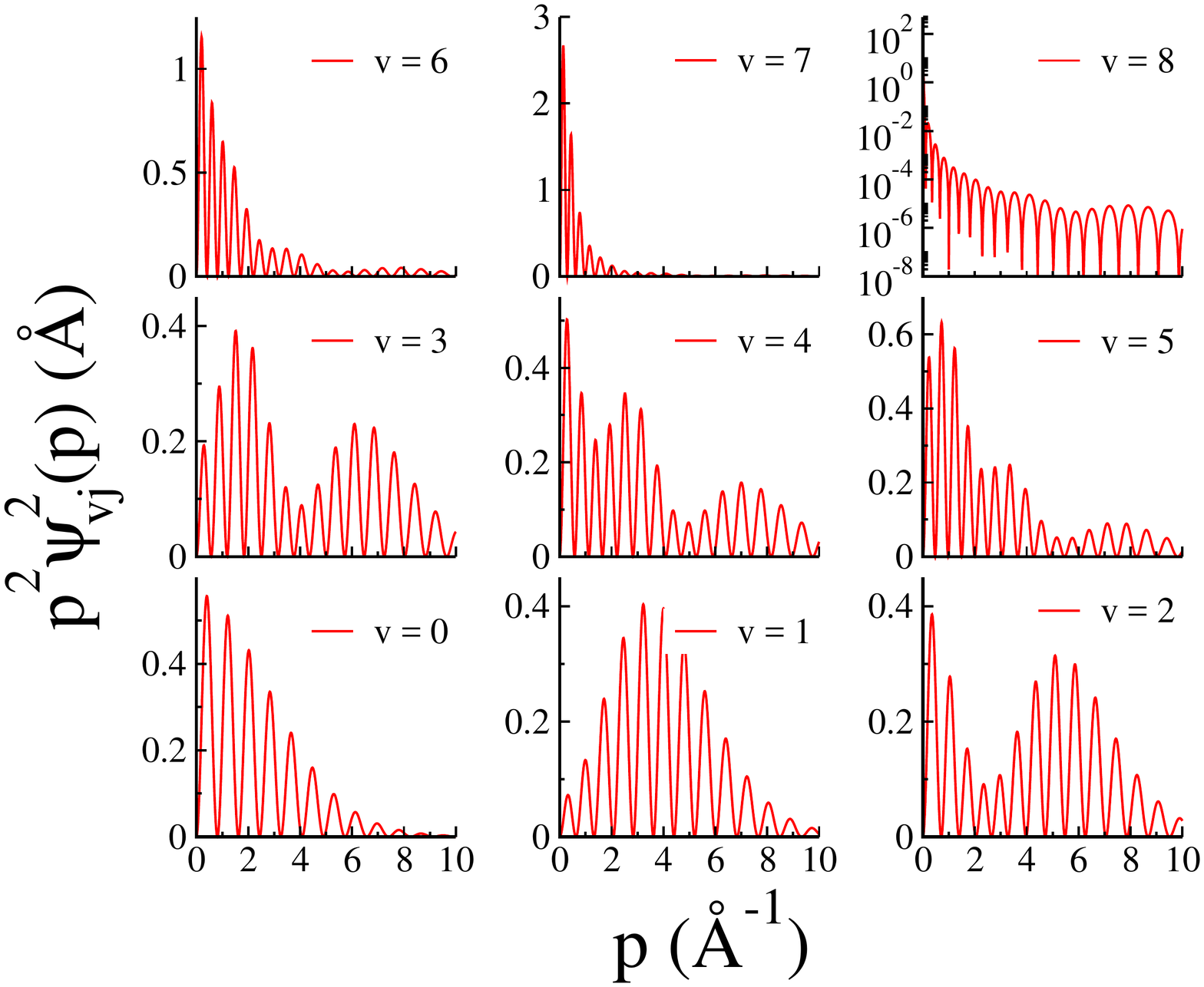}   \\
\includegraphics[width=.6\textwidth]{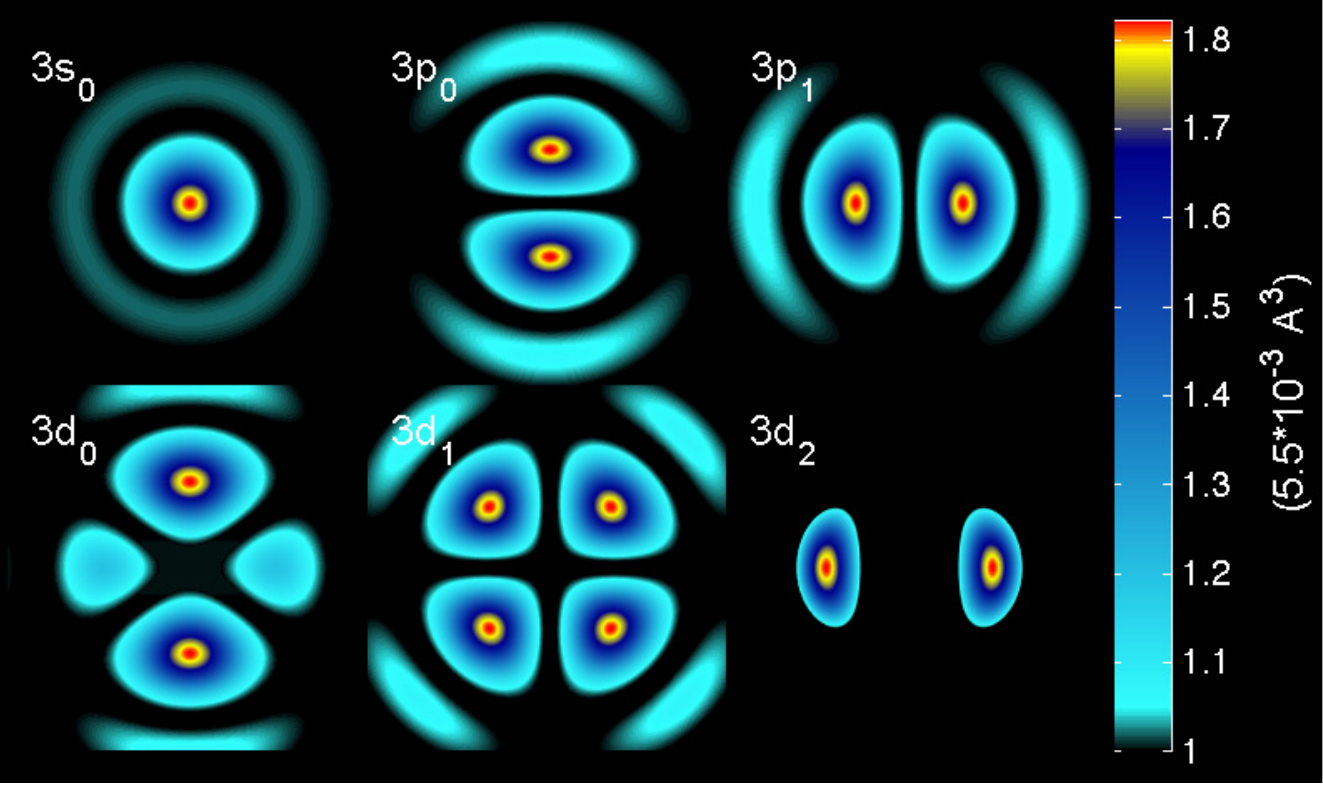}   \\
\end{array}  
$
\end{center}
\vspace{0.25in}
\hspace*{3in}
{\Large
\begin{minipage}[t]{3in}
\baselineskip = .5\baselineskip
Figure 3 \\
Taghi Sahraeian, M.~R.~Hadizadeh  \\
Int. J.\ Quant.\ Chem.
\end{minipage}
}

\clearpage

\begin{table}[hbt] 
\centering
\begin{tabular}{lcccc}
\hline
Parameter & Unit &  \multicolumn{2}{c}{Value}  \\  \cline{3-4} 
& & Model I \cite{Patkowski_MP103}   & Model II \cite{Patkowski_JCP133} \\ 
\hline
 $r_m$    & \AA  &  $0.529177209$     &  $0.529177209$ \\
$D$        & \AA   &  $0.15$  & $1.3$   \\
$A$         & cm$^{-1}$    &  $-29455669.7909$   &  $127641878.945519894$ \\
$B$         & cm$^{-1}$  &  $13812107.388$   &  $-26138949.621478189$     \\
$C$         & cm$^{-1}$  &  $45403994.0943$  & $-115672346.174201056$   \\
$D$         & cm$^{-1}$  &  $0$  &  $2064381.526204719$  \\
$E$         & cm$^{-1}$ &  $0$  &  $-58371.409016267$   \\
$\alpha$    &  - &  $1.623806026$ &  $1.553386357296$     \\
$\beta$      & - &  $0.0467301127$ &  $0$     \\
$b$        & - &  $1.500497187$ & $2.393847610341$         \\
$C_6$     & cm$^{-1}$     &  $64.691$  &  $64.288984$     \\
$C_8$      & cm$^{-1}$  &  $1644.0$  &  $1514.86211$     \\
$C_{10}$   & cm$^{-1}$    &  $50240.0$   &  $50240.0$     \\
$C_{12}$  & cm$^{-1}$   &  $1898195$   &  $1898195$      \\
$C_{14}$   & cm$^{-1}$     &  $86445426$  &  $86445426$    \\
$C_{16}$    & cm$^{-1}$   &  $4619452502$    &  $4619452502$     \\
$A'$         & cm$^{-1}$ &  $0$   &  $-383.344649817$  \\
$B'$          & cm$^{-1}$ &  $0$  & $152.167948910$    \\
$C'$        & cm$^{-1}$ &  $32837.87343$  & $324.0$     \\
$D'$         & cm$^{-1}$ &  $0$   & $-20.243797654$    \\
$\alpha'$   & -  &  $19.7726179$  & $2.3577$      \\
$\beta'$     & - &  $0$  &  $-1.2756$     \\
\hline
\end{tabular}
\caption{The parameters of the fitted analytical potential functions for argon-argon interaction given in Eq. (\ref{Vr}).}
\label{Table.Ar-parameters}
\end{table}

\begin{table}[hbt] 
\centering
\resizebox{\textwidth}{!}{%
\begin{tabular}{ccccccccccccccccc}
\hline
$j$           &  $v=0$ &   $v=1$   & $v=2$ & $v=3$  & $v=4$ & $v=5$ & $v=6$ & $v=7$ & $v=8$    \\ \hline
$0$ & $-84.38222$  & $-58.63217$   & $-38.14892$  & $-22.70572$ & $-11.91107$ & $-5.154345$ & $-1.594937$ & $-0.228064$  & $-0.86233 \cdot 10^{-6}$ \\
$1$ & $-84.26736$  & $-58.52605$  &  $-38.05267$ & $-22.62073$  & $-11.83897$ & $-5.096908$ & $-1.553999$ & $-0.205519$  \\
$2$ & $-84.03767$  & $-58.31384$  &  $-37.86022$ & $-22.45081$  & $-11.69486$ & $-4.982220$ & $-1.472475$ & $-0.161343$ \\
$3$ & $-83.69319$  & $-57.99561$   & $-37.57166$ & $-22.19612$ &  $-11.478982$ & $-4.810658$ & $-1.351087$ & $-0.097597$\\
$4$ & $-83.23399$  & $-57.57145$ &   $-37.18713$ & $-21.85686$ &  $-11.191667$ & $-4.582796$ & $-1.190966$ & $-0.018327$ \\
$5$ & $-82.66018$  & $-57.04150$ &   $-36.70682$ & $-21.43331$ &  $-10.833370$ & $-4.299421$ & $-0.993723$\\
$6$ & $-81.97188$  & $-56.40593$  &  $-36.13097$ & $-20.92584$ &  $-10.404669$ & $-3.961551$ & $-0.761562$ \\
$7$ & $-81.16922$  & $-55.66492$  &  $-35.45986$ & $-20.33488$ &  $-9.9062752$ & $-3.570465$ & $-0.497495$ \\
$8$ & $-80.25238$  & $-54.81872$  &  $-34.69385$ & $-19.66096$ &  $-9.3390434$ & $-3.127738$ & $-0.205754$  \\ 
$9$ & $-79.22155$  & $-53.86760$  &  $-33.83331$ & $-18.90468$ &  $-8.7039824$ & $-2.635303$ & $$   \\ 
$10$ & $-78.07696$  & $-52.81187$  & $-32.87870$& $-18.06673$ &  $-8.0022732$ & $-2.095521$ & $$   \\ 
$11$ & $-76.81886$  & $-51.65186$  & $-31.83050$& $-17.14792$ &  $-7.2352902$ & $-1.511309$ & $$    \\ 
$12$ & $-75.44751$  & $-50.38797$  & $-30.68934$& $-16.14914$ &  $-6.4046282$ & $-0.886329$ & $$    \\ 
$13$ & $-73.96322$  & $-49.02061$  & $-29.45580$& $-15.07141$ &  $-5.5121384$ & $-0.225320$ & $$    \\ 
$14$ & $-72.36631$  & $-47.55025$  & $-28.13059$& $-13.91588$ &  $-4.5599764$ & & $$    \\ 
$15$ & $-70.65715$  & $-45.97740$  & $-26.71448$& $-12.68383$ &  $-3.5506672$  \\ 
$16$ & $-68.83611$  & $-44.30260$  & $-25.20833$& $-11.37671$  &  $-2.4871984$   \\ 
$17$ & $-66.90363$  & $-42.52647$  & $-23.61307$& $-9.996130$ &  $-1.3731559$   \\ 
$18$ & $-64.86014$  & $-40.64966$  & $-21.92972$& $-8.543938$ &  $-0.2129375$   \\ 
$19$ & $-62.70613$  & $-38.67286$  & $-20.15943$& $-7.022209$    \\ 
$20$ & $-60.44210$  & $-36.59685$  & $-18.30342$& $-5.433312$    \\ 
$21$ & $-58.06861$  & $-34.42244$  & $-16.36306$& $-3.779966$    \\ 
$22$ & $-55.58623$  & $-32.15052$  & $-14.33987$& $-2.065334$    \\ 
$23$ & $-52.99560$  & $-29.78206$  & $-12.23550$& $-0.293139$    \\ 
$24$ & $-50.29738$  & $-27.31809$  & $-10.05181$   \\ 
$25$ & $-47.49226$  & $-24.75972$  & $-7.790866$   \\ 
$26$ & $-44.58100$  & $-22.10818$  & $-5.454975$   \\ 
$27$ & $-41.56440$  & $-19.36477$  & $-3.046762$   \\ 
$28$ & $-38.44330$  & $-16.53094$  & $-0.569217$   \\ 
$29$ & $-35.21860$  & $-13.60824$      \\ 
$30$ & $-31.89126$  & $-10.59837$      \\ 
$31$ & $-28.46232$  & $-7.503232$      \\ 
$32$ & $-24.93285$  & $-4.324887$      \\ 
$33$ & $-21.30403$  & $-1.065651$      \\ 
$34$ & $-17.57710$      \\ 
$35$ & $-13.75342$      \\ 
$36$ & $-9.834427$    \\ 
$37$ & $-5.821690$    \\ 
$38$ & $-1.716906$     \\ 
\hline
\end{tabular}
}
\caption{Argon dimer binding energies in units of cm$^{-1}$, calculated for potential model I.}
\label{Table.BE_spd1}  
\end{table}

\begin{table}[hbt] 
\centering
\resizebox{\textwidth}{!}{%
\begin{tabular}{ccccccccccccccccc}
\hline
$j$           &  $v=0$ &   $v=1$   & $v=2$ & $v=3$  & $v=4$ & $v=5$ & $v=6$ & $v=7$ & $v=8$    \\ \hline
$0$ & $-84.53458$  & $-58.85674$  & $-38.36106$ & $-22.85141$  & $-11.97942$ & $-5.171926$ & $-1.595389$ & $-0.227216$ & $-0.201861 \cdot 10^{-6}$ \\
$1$ & $-84.41941$  & $-58.75028$  & $-38.26443$ & $-22.76600$  & $-11.90692$ & $-5.114191$ & $-1.554288$ & $-0.204622$ & \\
$2$ & $-84.18909$  & $-58.53740$  & $-38.07122$ & $-22.59526$  & $-11.76203$ & $-4.998912$ & $-1.472442$ & $ -0.160358$ &  & \\
$3$ & $-83.84367$  & $-58.21815$  & $-37.78152$  & $-22.33933$  & $-11.544971$  & $-4.826465$  & $-1.350579$  & $-0.096505$   \\
$4$ & $-83.38322$  & $-57.79264$  & $-37.39546$  & $-21.99841$  & $-11.25608$  & $-4.597429$  & $-1.189841$  & $-0.017165$    \\
$5$ & $-82.80785$  & $-57.26100$  & $-36.91325$  & $-21.57280$  & $-10.89582$  & $-4.312598$  & $-0.991850$  & $$ \\
$6$ & $-82.11767$  & $-56.62340$  & $-36.33512$  & $-21.06285$  & $-10.46477$  & $-3.972996$  & $-0.758834$  & $$  \\
$7$ & $-81.31284$  & $-55.88004$  & $-35.66136$  & $-20.46900$  & $-9.963636$  & $-3.579912$  & $-0.493834$  & $$   \\
$8$ & $-80.39352$  & $-55.03116$  & $-34.89231$  & $-19.79178$  & $-9.393282$  & $-3.134934$  & $-0.201129$  & $$  \\ 
$9$ & $-79.35992$  & $-54.07702$  & $-34.02836$ & $-19.03178$  & $-8.754718$  & $-2.640006$  & $$  & $$   \\ 
$10$ & $-78.21225$  & $-53.01794$  & $-33.06996$ & $-18.18971$  & $-8.049131$  & $-2.097511$  & $$  & $$  \\
$11$ & $-76.95077$  & $-51.85427$  & $-32.01762$ & $-17.26636$  & $-7.277899$  & $-1.510388$  & $$  & $$  \\
$12$ & $-75.57576$  & $-50.58639$  & $-30.87191$ & $-16.26263$  & $-6.442622$  & $-0.882329$  & $$  & $$  \\
$13$ & $-74.08752$  & $-49.21472$  & $-29.63346$ & $-15.17955$  & $-5.545158$  & $-0.218119$  & $$  & $$  \\
$14$ & $-72.48639$  & $-47.73974$  & $-28.30298$ & $-14.01825$  & $-4.587669$  & $$  & $$  & $$  \\
$15$ & $-70.77273$  & $-46.16195$  & $-26.88122$ & $-12.78002$  & $-3.572690$  & $$  & $$  & $$  \\
$16$ & $-68.94693$  & $-44.48192$  & $-25.36905$ & $-11.46631$  & $-2.503218$  & $$  & $$  & $$  \\
$17$ & $-67.00942$  & $-42.70025$  & $-23.76741$ & $-10.07873$  & $-1.3828504$  & $$  & $$  & $$  \\
$18$ & $-64.96064$  & $-40.81759$  & $-22.07732$ & $-8.619129$  & $-0.2160005$  & $$  & $$  & $$  \\
$19$ & $-62.80109$  & $-38.83466$  & $-20.29990$ & $-7.089572$  & $$  & $$  & $$  & $$  \\
$20$ & $-60.53129$  & $-36.75223$  & $-18.43641$ & $-5.492429$  & $$  & $$  & $$  & $$  \\
$21$ & $-58.15179$  & $-34.57111$  & $-16.48821$ & $-3.830417$  & $$  & $$  & $$  & $$  \\
$22$ & $-55.66318$  & $-32.29221$  & $-14.45680$ & $-2.106689$  & $$  & $$  & $$  & $$  \\
$23$ & $-53.06610$  & $-29.91649$  & $-12.34385$ & $-0.324959$  & $$  & $$  & $$  & $$  \\
$24$ & $-50.36123$  & $-27.44499$  & $-10.1512$ & $$  & $$  & $$  & $$  & $$  \\
$25$ & $-47.54926$  & $-24.87883$  & $-7.88093$ & $$  & $$  & $$  & $$  & $$  \\
$26$ & $-44.63097$  & $-22.21925$  & $-5.535323$ & $$  & $$  & $$  & $$  & $$  \\
$27$ & $-41.60717$  & $-19.46755$  & $-3.117009$ & $$  & $$  & $$  & $$  & $$  \\
$28$ & $-38.47870$  & $-16.62518$  & $-0.628961$ & $$  & $$  & $$  & $$  & $$  \\
$29$ & $-35.24650$  & $-13.69370$  & $$ & $$  & $$  & $$  & $$  & $$  \\
$30$ & $-31.91152$  & $-10.67483$  & $$ & $$  & $$  & $$  & $$  & $$  \\
$31$ & $-28.47481$  & $-7.570446$  & $$ & $$  & $$  & $$  & $$  & $$  \\
$32$ & $-24.93747$  & $-4.382638$  & $$ & $$  & $$  & $$  & $$  & $$  \\
$33$ & $-21.30070$  & $-1.113719$  & $$ & $$  & $$  & $$  & $$  & $$  \\
$34$ & $-17.56574$  & $$  & $$ & $$  & $$  & $$  & $$  & $$  \\
$35$ & $-13.73398$  & $$  & $$ & $$  & $$  & $$  & $$  & $$  \\
$36$ & $-9.806879$  & $$  & $$ & $$  & $$  & $$  & $$  & $$  \\
$37$ & $-5.786016$  & $$  & $$ & $$  & $$  & $$  & $$  & $$  \\
$38$ & $-1.673116$  & $$  & $$ & $$  & $$  & $$  & $$  & $$  \\
\hline
\end{tabular}
}
\caption{Argon dimer binding energies in units of cm$^{-1}$, calculated for potential model II.}
\label{Table.BE_spd2}  
\end{table}

\begin{table}[hbt] 
\centering
\begin{threeparttable}[b]
\begin{tabular}{ccccccccccccccccc}
\hline
State &  \multicolumn{5}{c}{Model I \cite{Patkowski_MP103}} &  Exp. \cite{Herman_JCP89}  \\  \cline{2-6} 
    &   Present     &  Ref. \cite{Patkowski_private} &   $\Delta (\%)$   & R-matrix \cite{Tennyson_FD} &   $\Delta (\%)$   \\ \hline
$v=0$ & $-84.38222$ & $-84.38263$ &  $0.000486$ & $-84.38263$   & $0.000486$    &  $-84.47$  \\
$v=1$ & $-58.63217$ & $-58.63364$ &  $0.002507$ & $-58.63364$  & $0.002507$     &   $-58.78 \pm 0.01$ \\
$v=2$ & $-38.14892$ & $-38.15162$ &  $0.007077$ & $-38.15162$ &  $0.007077$     & $-38.2 \pm 0.02$   \\
$v=3$ & $-22.70572$ & $-22.70911$ &  $0.014928$ & $-22.70911$ &  $0.014928$     & $-22.62 \pm 0.02$   \\
$v=4$ & $-11.91107$ & $-11.91426$ &  $0.026775$ & $-11.91426$ &   $0.026775$    &  $-11.71 \pm 0.03$  \\
$v=5$ & $-5.15435$ & $-5.15662$ &     $0.044021$ & $-5.15662$ &   $0.044021$    &  $-4.87 \pm 0.07$   \\
$v=6$ & $-1.59494$ & $-1.59610$ &     $0.072677$ & $-1.59611$ &   $0.073303$     & $-$  \\
$v=7$ & $-0.22806$ & $-0.22840$ &     $0.148862$ & $-0.22840$ &   $0.148862$     &  $-$ \\
$v=8$ & $-0.86233 \cdot 10^{-6}$  & $-0.00000$ \tnote{1} &  $-$ & $-$ &  $-$ &    $-$   \\
\hline
State &  \multicolumn{5}{c}{Model II \cite{Patkowski_JCP133}} &  Exp. \cite{Herman_JCP89}  \\  \cline{2-6} 
    &   Present     &  Ref. \cite{Patkowski_private} &   $\Delta (\%)$ & R-matrix \cite{Rivlin_private} &   $\Delta (\%)$   \\ \hline
$v=0$ & $-84.53458$ & $-84.53499$   & $0.000485$ &   $-84.53495$  &   $0.000438$    & $-84.47$     \\
$v=1$ & $-58.85674$ & $-58.85820 $  & $0.002480$ &   $-58.85817$  &   $0.002430$     & $-58.78 \pm 0.01$   \\
$v=2$ & $-38.36106$ & $-38.36374$   & $0.006986$ &   $-38.36372$  &   $0.006934$     & $-38.2 \pm 0.02$   \\
$v=3$ & $-22.85141$ & $-22.85481$   & $0.014876$ &   $-22.85479$  &   $0.014789$    &  $-22.62 \pm 0.02$   \\
$v=4$ & $-11.97942$ & $-11.98263$   & $0.026789$ &   $-11.98262$  &    $0.026705$    &  $-11.71 \pm 0.03$   \\
$v=5$ & $-5.17193$ & $-5.17421$   &    $0.044065$ &   $-5.17420$  &       $0.043871$     &  $-4.87 \pm 0.07$  \\
$v=6$ & $-1.59539$ & $-1.59656$   &    $0.073283$ &   $-1.59656$  &      $0.073283$     &   $-$ \\
$v=7$ & $-0.22722$ & $-0.22755$   &    $0.145023$ &   $-0.22755$  &    $0.145023$    &    $-$ \\
$v=8$ & $-0.20186 \cdot 10^{-6}$  & $-0.00000$ \tnote{1}  & $-$  &        $-$  &   $-$     &   $-$ \\
\hline
\end{tabular}
\caption{The vibrational energy levels of the ground state of argon dimer, i.e. $j=0$, in units of cm$^{-1}$ calculated for potential models I and II, in comparison to the findings of other groups.
$\Delta$ is the absolute value of the relative percentage difference between our findings and the predictions of Ref. \cite{Patkowski_private} and R-matrix method \cite{Tennyson_FD, Rivlin_private}.}
\label{Table.BE_s}
\begin{tablenotes}
            \item [1] It is the maximum accuracy provided in Ref. \cite{Patkowski_private} for the ninth vibrational state.
        \end{tablenotes}
    \end{threeparttable}        
\end{table}

\begin{table}[hbt] 
\centering
\begin{threeparttable}[b]
\begin{tabular}{ccccccccccccccccc}
\hline
State &  \multicolumn{7}{c}{Model I \cite{Patkowski_MP103}}  \\  \cline{2-8} 
   &  \multicolumn{3}{c}{$j=1$} & \multicolumn{3}{c}{$j=2$}  \\  \cline{2-4}  \cline{6-8} 
    &   Present    & R-matrix \cite{Rivlin_private} &   $\Delta (\%)$ &  & Present    & R-matrix \cite{Rivlin_private} &   $\Delta (\%)$   \\ \hline
$v=0$ & $-84.26736$ & $-84.26778$ &  $0.000498$  && $-84.03767$ & $-84.03809$ &  $0.000500$   \\
$v=1$ & $-58.52605$ & $-58.52753$ &  $0.002529$  && $-58.31384$ & $-58.31534$ &  $0.002572$      \\
$v=2$ & $-38.05267$ & $-38.05538$ &  $0.007121$  && $-37.86022$ & $-37.86296$ &  $0.007237$      \\
$v=3$ & $-22.62073$ & $-22.62413$ &  $0.015028$  && $-22.45081$ & $-22.45425$ &  $0.015320$      \\
$v=4$ & $-11.83897$ & $-11.84216$ &  $0.026938$  && $-11.69486$ & $-11.69807$ &  $0.027440$      \\
$v=5$ & $-5.096908$ & $-5.099183$ &  $0.044615$  && $-4.982220$ & $-4.984496$ &  $0.045662$      \\
$v=6$ & $-1.553999$ & $-1.555162$ &  $0.074783$  && $-1.472475$ & $-1.473627$ &  $0.078174$      \\
$v=7$ & $-0.205519$ & $-0.205846$ &  $0.158857$  && $-0.161343$ & $-0.161651$ &  $0.190534$      \\
\hline
State &  \multicolumn{7}{c}{Model II \cite{Patkowski_JCP133}}  \\  \cline{2-8} 
   &  \multicolumn{3}{c}{$j=1$} & \multicolumn{3}{c}{$j=2$}  \\  \cline{2-4}  \cline{6-8} 
    &   Present    & R-matrix \cite{Rivlin_private} &   $\Delta (\%)$ &  & Present    & R-matrix \cite{Rivlin_private} &   $\Delta (\%)$   \\ \hline
    $v=0$ & $-84.41941$ & $-84.41978$ &  $0.000438$  && $-84.18909$ & $-84.18947$ &  $0.000451$   \\
$v=1$ & $-58.75028$ & $-58.75171$ &  $0.002434$  && $-58.53740$ & $-58.53885$ &  $0.002477$      \\
$v=2$ & $-38.26443$ & $-38.26711$ &  $0.007003$  && $-38.07122$ & $-38.07392$ &  $0.007091$      \\
$v=3$ & $-22.76600$ & $-22.76940$ &  $0.014932$  && $-22.59526$ & $-22.59868$ &  $0.015134$      \\
$v=4$ & $-11.90692$ & $-11.91012$ &  $0.026868$  && $-11.76203$ & $-11.76525$ &  $0.027369$      \\
$v=5$ & $-5.114191$ & $-5.116472$ &  $0.044581$  && $-4.998912$ & $-5.001193$ &  $0.045609$      \\
$v=6$ & $-1.554288$ & $-1.555451$ &  $0.074769$  && $-1.472442$ & $-1.473594$ &  $0.078176$      \\
$v=7$ & $-0.204622$ & $-0.204947$ &  $0.158578$  && $-0.160358$ & $-0.160663$ &  $0.189838$      \\
\hline
\end{tabular}
\caption{The vibrational energy levels of argon dimer for $j=1$ and $j=2$, in units of cm$^{-1}$ calculated for potential models I and II, in comparison to the findings of R-matrix method \cite{Rivlin_private}.
$\Delta$ is the absolute value of the relative percentage difference between our findings and R-matrix predictions.}
\label{Table.BE_pd}
    \end{threeparttable}        
\end{table}

\begin{table}[hbt] 
\centering
\begin{tabular}{cccccccccccc}
\hline
State &  \multicolumn{3}{c}{Model I}  \\  \cline{2-4} 
    &   Present     & Ref. \cite{Patkowski_MP103}    \\ \hline
$v=8$ & $-1.24069$  & $-0.74$ &  &  \\
\hline
\end{tabular}
\caption{The ninth vibrational energy level of the ground state of argon dimer, i.e. $j=0$, in units of $\mu$K calculated for the potential model I, in comparison to the finding of Ref. \cite{Patkowski_MP103}.
}
\label{Table.BE9}
\end{table}

\end{document}